\documentclass[twocolumn]{aastex63}
\usepackage{amsmath,amstext}
\usepackage{balance}
\usepackage{color}
\usepackage{comment}

\graphicspath{{./}{figures/}}
\begin{document}

\title{The X-ray pulsar XTE J1858+034 observed with NuSTAR and Fermi/GBM:\\
spectral and timing characterization plus a cyclotron line}

\correspondingauthor{cmalacaria@usra.edu}


\author[0000-0002-0380-0041]{C.~Malacaria}
\affiliation{NASA Marshall Space Flight Center, NSSTC, 320 Sparkman Drive, Huntsville, AL 35805, USA}\thanks{NASA Postdoctoral Fellow}
\affiliation{Universities Space Research Association, Science and Technology Institute, 320 Sparkman Drive, Huntsville, AL 35805, USA}

\author[0000-0001-9840-2048]{P.~Kretschmar}
\affiliation{European Space Agency (ESA), European Space Astronomy Centre (ESAC), Camino Bajo del Castillo s/n, 28692 Villanueva de la Cañada, Madrid, Spain}

\author[0000-0003-1252-4891]{K.K.~Madsen}
\affiliation{CRESST and X-ray Astrophysics Laboratory, NASA Goddard Space Flight Center, Greenbelt, MD 20771, USA}
\affiliation{Department of Physics and Center for Space Science and Technology, University of Maryland, Baltimore County, Baltimore, MD
21250, USA}

\author[0000-0002-8585-0084]{C.A.~Wilson-Hodge}
\affiliation{ST 12 Astrophysics Branch, NASA Marshall Space Flight Center, Huntsville, AL 35812, USA}

\author[0000-0001-7532-8359]{Joel~B.~Coley}
\affiliation{Department of Physics and Astronomy, Howard University, Washington, DC 20059, USA}
\affiliation{CRESST and Astroparticle Physics Laboratory, NASA Goddard Space Flight Center, Greenbelt, MD 20771, USA}

\author{P. Jenke}
\affiliation{University of Alabama in Huntsville (UAH), Center for Space Plasma and Aeronomic Research (CSPAR), 301 Sparkman Drive, Huntsville, Alabama 35899}

\author[0000-0002-6255-9972]{Alexander~A.~Lutovinov}
\affiliation{Space Research Institute of the Russian Academy of Sciences, Profsoyuznaya Str. 84/32, Moscow 117997, Russia}

\author[0000-0002-4656-6881]{K.~Pottschmidt}
\affiliation{CRESST and Astroparticle Physics Laboratory, NASA Goddard Space Flight Center, Greenbelt, MD 20771, USA}
\affiliation{Department of Physics and Center for Space Science and Technology, University of Maryland, Baltimore County, Baltimore, MD
21250, USA}

\author[0000-0002-9679-0793]{Sergey~S.~Tsygankov}
\affiliation{Department of Physics and Astronomy, FI-20014 University of Turku,  Finland}
\affiliation{Space Research Institute of the Russian Academy of Sciences, Profsoyuznaya Str. 84/32, Moscow 117997, Russia}

\author[0000-0003-2065-5410]{J.~Wilms}
\affiliation{Remeis-Observatory and Erlangen Centre for Astroparticle Physics, Friedrich-Alexander-Universit\"at Erlangen-N\"urnberg, Sternwartstr.~7, 96049 Bamberg, Germany}

\begin{abstract}
Accreting X-ray pulsars (XRPs) undergo luminous X-ray outbursts during which the spectral and timing behavior of the neutron star can be studied in detail. We analyze a \textit{NuSTAR} observation of the XRP XTE J1858+034 during its outburst in 2019.
The spectrum is fit with a phenomenological, a semi-empirical and a physical spectral model.
A candidate cyclotron line is found at $48\,$keV, implying a magnetic field of $5.4\times10^{12}\,$G at the site of emission. This is also supported by the physical best-fit model.
We propose an orbital period of about 81 days based on the visual inspection of the X-ray outbursts recurrence time. Based on \textit{Fermi} Gamma-ray Burst Monitor data, the standard disk accretion-torque theory allowed us to infer a distance of $10.9\pm1.0\,$kpc.
Pulse profiles are single-peaked and show a pulsed fraction that is strongly energy-dependent at least up to 40\,keV.
\end{abstract}

\keywords{X-ray binary stars -- stars: neutron -- pulsars: individual: XTE J1858+034 -- accretion, accretion disks -- magnetic fields}

\section{Introduction}\label{sec:introduction}

Accreting X-ray pulsars (XRPs) are binary systems consisting of a neutron star (NS) that accretes matter originating from a donor companion star
via stellar wind or Roche-lobe overflow.
XTE J1858+034 is an XRP discovered with the \textsl{Rossi X-ray Timing Explorer (RXTE)} in 1998 by \citet{Remillard98} and \citet{Takeshima98}. Those observations also detected X-ray pulsations with a period of ${\sim}221\,$s. 
X-ray emission from this source has been detected only in a few short outbursts \citep[and references therein]{Nakajima19}, thus preventing to obtain an orbital solution or an in-depth characterization of the system.
A Cyclotron Resonant Scattering Feature (CRSF) also was not observed from this source so far.
When observed, the energy $E_\mathrm{cyc}$ of the fundamental CRSF probes the magnetic field strength at the site of spectral emission,  $E_\mathrm{cyc}\sim11.6{\times}B_{12} (1 + z_\mathrm{g})^{-1}\,$keV, where $B_{12}$ is the magnetic field in units of $10^{12}\,$G, and $z_\mathrm{g}$ is the gravitational redshift (see \citealt{Staubert19} for a recent review).
However, \citet{Paul98} estimated a magnetic field strength of $0.8\times10^{12}\times d_{kpc}\,$G (with $d_{kpc}$ the distance value in units of kpc), based on the observation of quasi-periodical oscillations in this system.

\citet{Reig04atel, Reig05} proposed a Be-type star for the optical counterpart, of which neither the spectral subtype nor the distance was found. This star was the only one within the hard X-ray error circle from \textit{INTEGRAL} observations \citep{Molkov04} showing H$\alpha$ emission and was thus proposed as counterpart although it lay outside the error circle of the JEM-X soft X-ray instrument.
At an angular offset of $3.5\arcsec$ from the nominal X-ray source position, \textit{Gaia} found an optical candidate counterpart \citep{Bailer-Jones18}, mentioned in \citet{Malacaria20} as possible counterpart, but likely unassociated given the large offset.
In addition, the \textit{Gaia} counterpart is at an angular offset of $103\arcsec$ from, and thus clearly not associated with, the optical counterpart proposed by \citet{Reig05}. This question is discussed in more detail in the accompanying paper by Tsygankov et al. (in press), who identify a counterpart based on \textit{Chandra} and ground telescope observations, with a probable distance of 7--14~kpc.


\begin{figure}[!t]
\includegraphics[width=.48\textwidth]{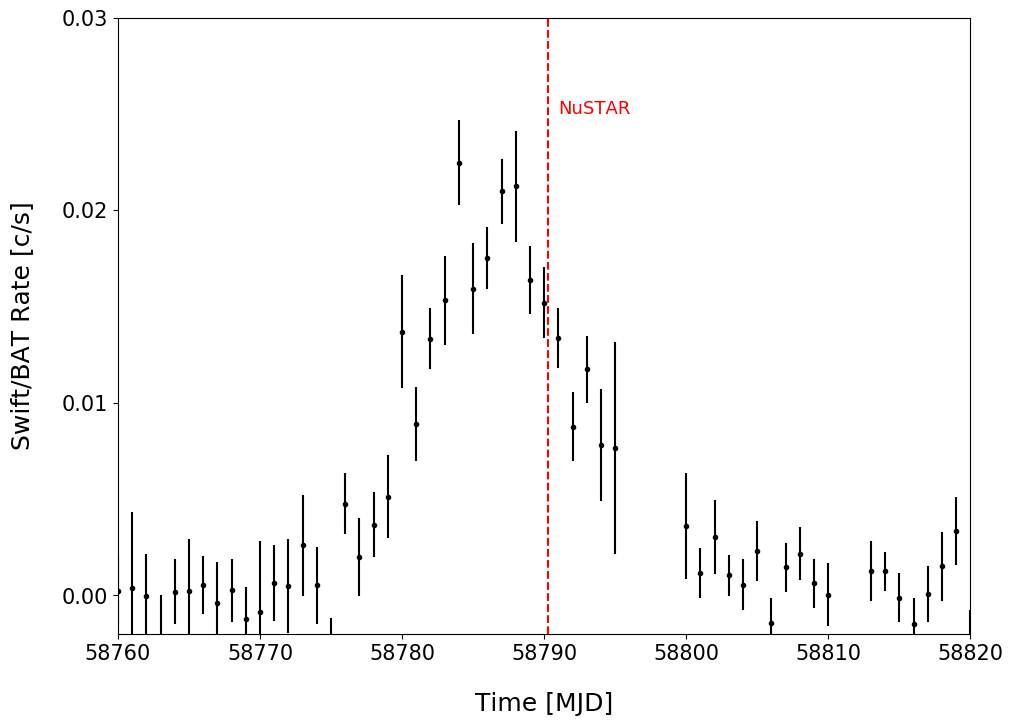}
\caption{\textsl{Swift}/BAT daily average light curve of XTE J1858+034 during the outburst in 2019 (black dots). The \textit{NuSTAR} observation time is also shown (red dashed line). \label{fig:outburst}}
\end{figure}

\begin{figure*}[!t]
\centering
\includegraphics[width=.95\textwidth]{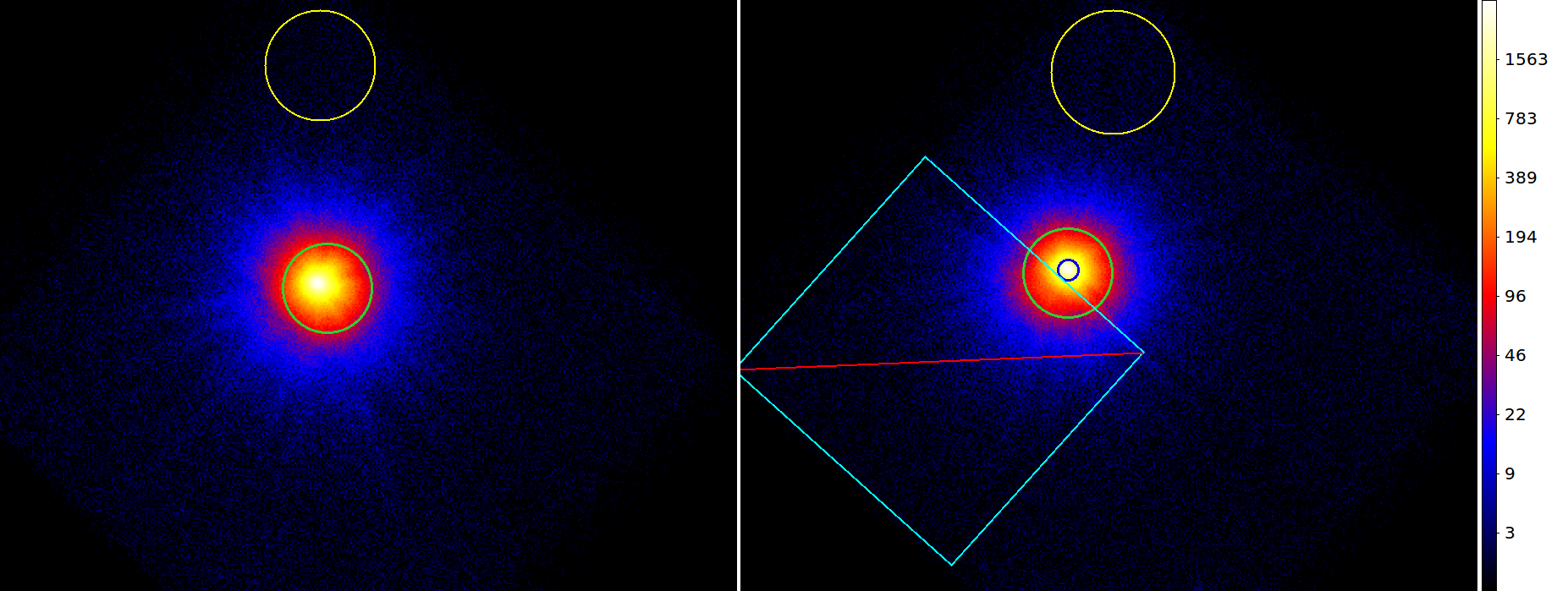}
\caption{\textit{NuSTAR} images of XTE J1858+034 as observed in November 2019 from FPMA (left) and FPMB (right). Circular green regions centered on the source represent the source extraction regions. Yellow solid circles on the top corner represent the background extraction regions. For FPMB, the blue small circle represents the ARF extraction region, while the barred cyan square represents the exclusion of detector 3 (see text). The color bar shows the number of counts per pixel.\label{fig:1}}
\end{figure*}

Recently, the source has undergone a new outburst episode \citep{Nakajima19}, and was observed with \textit{NuSTAR}.
Here we study its spectral and timing characteristics and finally form a consistent general overview for the X-ray behaviour of XTE J1858+034, including a distance estimate based on accretion-torque theory. 
The analysis presented here is complemented by the work in the accompanying paper by Tsygankov et al. (in press).


\section{Data reduction}\label{sec:data_reduction}

\textit{NuSTAR} \citep{Harrison13} was launched in 2012. It is currently the only X-ray mission with a telescope able to focus hard X-rays above 10 keV.
\textit{NuSTAR} consists of two identical co-aligned telescopes that focus X-ray photons onto two independent Focal Plane Modules, FPMA and FPMB.
At the focus of each telescope module are four ($2\times2$) solid-state cadmium zinc telluride (CdZnTe) imaging detectors.
These provide wide-band (3--79\,keV) energy coverage with a FWHM of $18\arcsec$ and a spectral resolution of 400\,eV at 10\,keV. 

\textit{NuSTAR} observed XTE J1858+034 on 2019 November 3 (ObsID 90501348002, MJD 58790), during an outburst (see Fig.~\ref{fig:outburst}).
The total exposure time was about 44 ks.
\textit{NuSTAR} data were reduced with \texttt{NUSTARDAS} v1.9.5 provided by the \texttt{HEASOFT} v6.27.2 and using the \textit{CALDB} 20200526 \citep{Madsen20}.
Cleaned events were obtained following the standard \textit{NuSTAR} guidelines.
The resulting images are shown in Fig.~\ref{fig:1}.
Source spectra were extracted through the \texttt{NUPRODUCTS} routine.
The source extraction region was a $65\arcsec$ radius circular region centered on the source, while the background was extracted from a source-free region on the same detector with radii of $90\arcsec$ and $105\arcsec$ for FPMA and FPMB, respectively.
We also verified that shifting the extraction regions in order to account for the offset between the images from the two modules does not significantly affect the results.
However, in FPMB part of the source events fall on the chip gap between detectors 0 and 3, resulting in unaccounted loss of effective area. 
Moreover, \textit{NuSTAR} detectors suffer from absorption due to a CdZnTe dead layer and a Pt coating at the top of the detectors, an effect that is calibrated through observations of the Crab in stray light mode \citep{Madsen_2017}. 
However, the absorption curve can be degenerate with other effective area effects, and for detector 3 this has caused part of the detector-related absorption to be included in the vignetting curve, thus resulting in spectral differences in the low energy spectrum when compared to detector 0 (priv. comm. with the \textit{NuSTAR} Science Operations Team).
All these factors led us to exclude the entire detector 3 from the FPMB source extraction region. 
We obtained the FPMB ARF from a $15\arcsec$ radius circular region centered on the source, which ensures that the detector absorption of detector 3 did not get included.
Similarly, to avoid accidentally including the RMF from detector 3 during RMF generation the RMF for detector 0 was obtained directly from the CALDB (\texttt{nuBcutdet0\_20100101v001}).

Spectral data were analyzed using \texttt{XSPEC} v12.11.0l \citep{Arnaud96}.
\textit{NuSTAR} data were used in the range $3-60\,$keV ($3.5-60\,$keV for FPMB to further enhance consistency between spectra in the lowest channels), above which the background dominates.
Spectra were rebinned to have at least $50$ counts per bin.

\section{Results}
\subsection{Spectral analysis}\label{subsec:spectral}

XTE J1858+034 as observed by \textit{NuSTAR} in November 2019 clearly shows a hard spectrum.
FPMA and FPMB spectra have been fitted simultaneously, allowing for a cross-normalization factor.
Although the cross-normalization factor between FPMA and FPMB is usually of the order of a few percent \citep{Madsen15}, the limited ARF extraction region adopted in our analysis for FPMB (see Sect.~\ref{sec:data_reduction}) is expected to reduce the cross-normalization value significantly.
For the spectral fit, standard phenomenological and semi-empirical continuum models have been employed, namely two variants of the cutoff power-law model (\texttt{cutoffpl} and \texttt{highecut*pow} in \texttt{XSPEC}) and a Comptonization model of soft photons in a hot plasma (\texttt{compTT} in \texttt{XSPEC}, \citealt{Titarchuk94}), respectively.
To obtain an acceptable fit, the \texttt{cutoffpl} and \texttt{highecut*pow} models need an additional component in the lower energy band, which has been modeled as a blackbody emission as found in other accreting XRPs (see, e.g., \citealt{Palombara06}).
However, the blackbody temperature is high with respect to other XRPs, indicating that the phenomenological model is likely inadequate.
Moreover, we also tested a purely-physical model of thermal and bulk Comptonization of the seed photons produced by cyclotron cooling \citep[\texttt{bwcycl} in  \texttt{XSPEC}]{Ferrigno09}.
For a fixed value of mass and radius of the accreting NS, the \texttt{bwcycl} model has six free parameters, namely the accretion rate $\dot{M}$, the magnetic field strength $B$, the accretion column radius $r_0$, the electron temperature $T_e$, the photon diffusion parameter $\xi$ and the Comptonization parameter $\delta$.
This model was successfully used to fit the broad-band energy spectrum of a number of bright ($\gtrsim10^{37}\,$erg\,s$^{-1}$) accreting XRPs (see, e.g., \citealt{Epili19,Dai17,Wolff16}).

For all tested models, the photoelectric absorption component and elemental abundances were set according to \citealt{Wilms00} (\texttt{tbabs} in  \texttt{XSPEC}) to account for photoelectric absorption by neutral interstellar matter (or column density $N_H$), and assuming model-relative (\texttt{wilm}) solar abundances.
Given that the Galactic $N_H$ in the direction of the source is about $1.7\times10^{22}\,$cm$^{-2}$ \citep{HI4PI2016}, all models show important local absorption values.
All tested models also were equipped with a Gaussian emission line at 6.4 keV to account for the Fe K$\alpha$ fluorescence emission.

\begin{figure}[!t]
\includegraphics[width=0.48\textwidth]{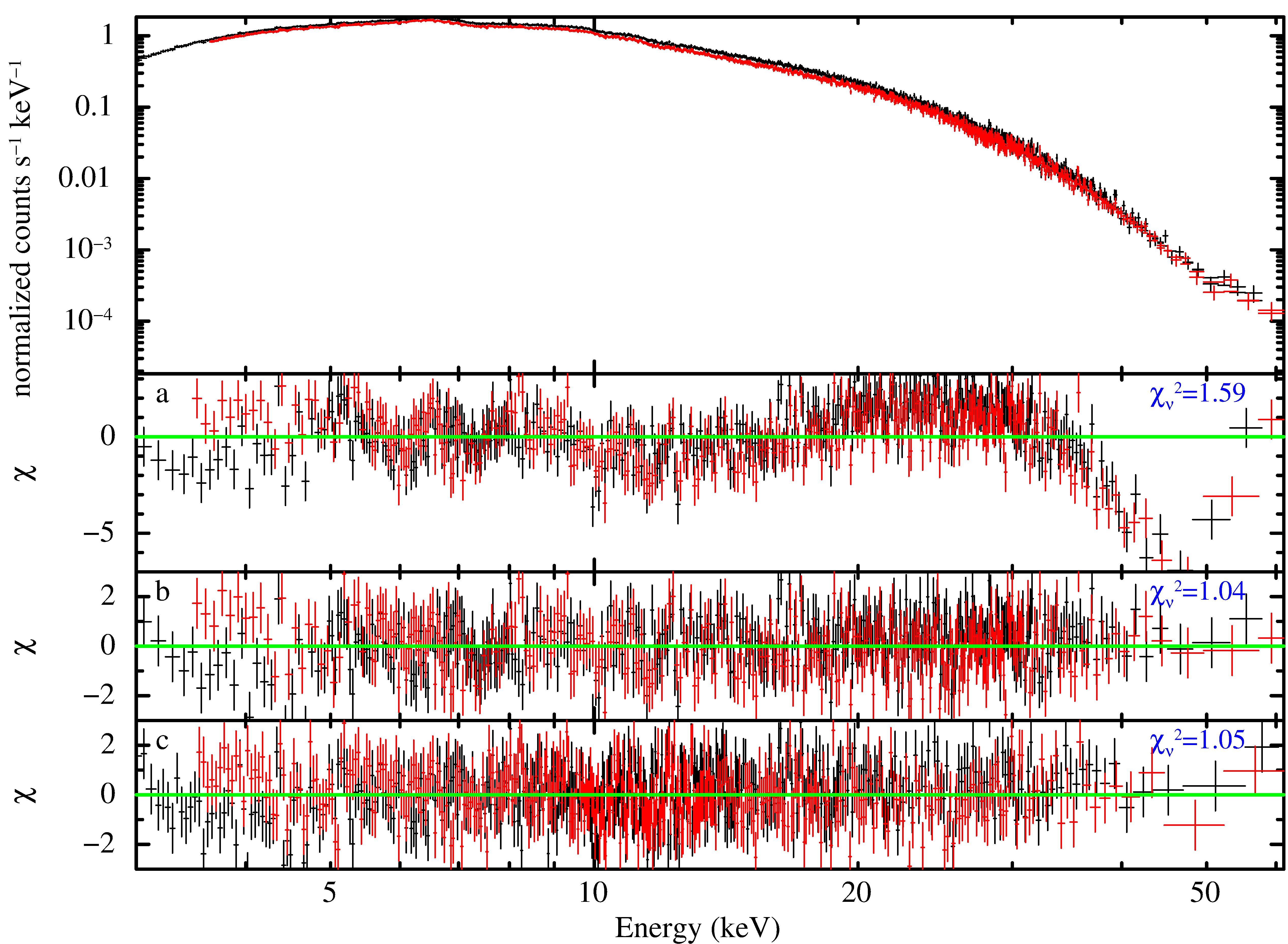}
\caption{\textit{Top}: XTE J1858+034 spectrum as observed by \textit{NuSTAR} in 2019 and fit with a \texttt{ComptTT} model. Lower panels are referred with a letter in the upper left corner.
\textit{Panel a}: residuals of the \texttt{ComptTT} model.
\textit{Panel b}: residuals of the best-fit \texttt{ComptTT} model including a Gaussian absorption line at $\sim48\,$keV (see Table~\ref{table:spectral}).
\textit{Panel c}: residuals of the best-fit \texttt{BWCYC IIa} model including a Gaussian absorption line at $\sim48\,$keV (see Table~\ref{table:spectral2}).
Spectra and residuals have been rebinned for plotting purpose.
The blue text in the right corners of the lower panels shows the correspondent model $\chi^2$ divided by $\nu$ degrees of freedom.
\label{fig:spec}}
\end{figure}

\begin{figure}[!t]
\includegraphics[width=0.49\textwidth]{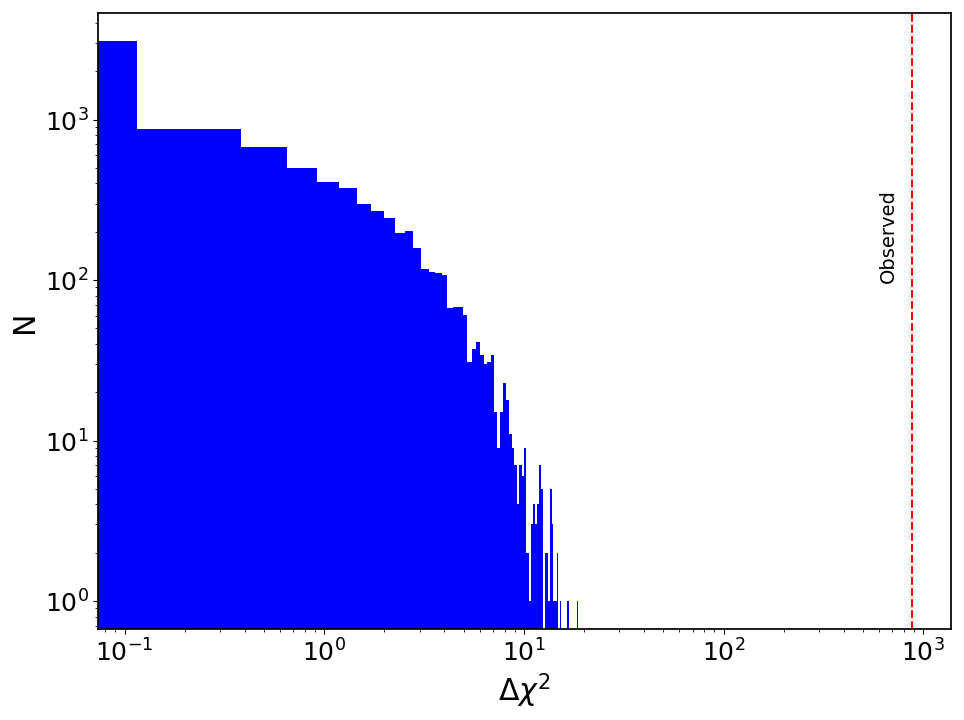}
\caption{Results of 10000 Monte Carlo simulations to test the significance of the Gaussian absorption line in the \texttt{CompTT} model.
The solid histogram shows the frequency (y-axis) of $\Delta\chi^2$ values (x-axis) obtained in the simulation.
The red dashed line shows the observed $\Delta\chi^2=878$.
\label{fig:histo}}
\end{figure}

All fit continuum models show absorption-like residuals in the range $40-50\,$keV.
These residuals can be modeled with a Gaussian absorption line (see Fig.~\ref{fig:spec}).
The improvement in the best-fit statistics is maximum in the \texttt{compTT} model, i.e. $\Delta\chi^2=878$.
Other models show an improvement of $\Delta\chi^2\gtrapprox400$, with the lowest $\Delta\chi^2$ derived from the \texttt{highecut*pow} model.
The significance of the line in the \texttt{compTT} model has been assessed through Monte Carlo simulations.
For this task, the \texttt{XSPEC simftest} routine was adopted, which allows to simulate a chosen number of spectra based on the actual data and test the resulting $\Delta\chi^2$ between each instance fit when the additional model component (the Gaussian absorption line in our case) is included.
Following \citet{Bhalerao15, Bodaghee16}, the column density parameter was fixed to its best-fit value, and the energy and width of the Gaussian absorption line were left free to vary within their $90\%$ confidence region in order to improve the speed and convergence of the fits.
Simulations results are reported in Fig.~\ref{fig:histo} for a $10^4$ iterations process and confirm the significance of the absorption feature at $>3\sigma\,$c.l.
Following \citet{Marcu+15}, we also investigated the impact of a variable background normalization on the absorption feature parameters.
Using the \texttt{XSPEC} tool \texttt{recorn}, it was found that the absorption line parameters do not change significantly if the normalization of the background spectrum is increased up to a $50\%$ higher level, thus strengthening the interpretation of the absorption feature as real and not due to artifacts.
The feature was also observed in phase-resolved spectra presented in the accompanying paper by Tsygankov et al. (in press).
Interpreting the feature at $48\,$keV as a CRSF, and assuming a gravitational redshift of $z_g=0.3$ (for NS mass and radius of 1.4\,$M_\odot$ and 10 km, respectively), a magnetic field strength of B$=(5.4\pm0.1)\times10^{12}\,$G is obtained.

Following the \texttt{bwcycl} model instructions\footnote{\url{https://heasarc.gsfc.nasa.gov/xanadu/xspec/manual/node148.html}.}, it is convenient to freeze some of the model parameters in order to improve the computational speed and help the fit converge to the best-fit parameters.
Once the best-fit was found, the column density $N_H$ also was fixed to its best-fit value to help the fit converge and to obtain parameters errors.
Mass and radius of the NS were fixed to their canonical values of $1.4$M$_\odot$ and $10\,$km, respectively.
However, it is preferable to also fix the values of the NS magnetic field, its distance and its mass accretion rate (as derived by the observed luminosity).
For XTE J1858+034, there are no previous conclusive estimations of the magnetic field, while a measurement of the distance is necessary for the latter two parameters.
As mentioned in Sect.~\ref{sec:introduction},
the closest \textit{Gaia} counterpart to the nominal X-ray position found by \citet{Molkov04} was unlikely associated with the X-ray source or with the optical counterpart proposed by \citet{Reig05}.
This was ascertained by Tsygankov et al. (in press), who shown that either of those possible counterparts is consistent with the much better constrained X-ray source location available through new \textit{Chandra} observations.
However, the distance to the X-ray system had not been estimated before their work, which in any case did not constrain it very much.
Therefore, we opted for a different approach to obtain a more stringent value of the distance, based on the spin-up ($\dot{P}$) measured by Fermi-GBM.

\begin{table}[!t]
\caption{Best-fit results of XTE J1858+034 spectral analysis with a cutoff power-law model \texttt{cutoffpl} and a Comptonization model \texttt{CompTT}. All reported errors are at $1\sigma\,$c.l.} \label{table:spectral}
\begin{ruledtabular}\begin{tabular}{lcc}
 & cutoffpl & CompTT \\
  & \\
C$_{FPMB}$ &$0.747^{+0.001}_{-0.001}$ & $0.747^{+0.001}_{-0.001}$ \\
N$_{\textrm{H}}$ [$10^{22}\,$cm$^{-2}$] & $7.6^{+0.4}_{-0.4}$ & $5.8^{+0.3}_{-0.3}$\\
kT$_{bb}\,$[keV] & $5.2^{+0.2}_{-0.1}$ & -- \\
norm$_{bb}$ & $0.0152^{+0.0002}_{-0.0007}$ & -- \\
E$_{K\alpha}\,$[keV] & $6.47^{+0.02}_{-0.02}$ & $6.48^{+0.02}_{-0.02}$ \\
$\sigma_{K\alpha}\,$[keV] & $0.26^{+0.02}_{-0.02}$ & $0.28^{+0.02}_{-0.02}$ \\
norm$_{K\alpha}$ ($10^{-4}$) & $5.5^{+0.4}_{-0.4}$ & $5.9^{+0.4}_{-0.4}$ \\
$\Gamma$ & $0.03^{+0.29}_{-0.22}$ & -- \\
HighECut [keV] & $3.5^{+1.5}_{-0.5}$ & -- \\
norm$_{\Gamma}$$^*$ & $0.025^{+0.006}_{-0.004}$ & -- \\
T0 [keV] & -- & $1.02^{+0.02}_{-0.02}$ \\
kT$_{CompTT}$ [keV] & -- & $5.61^{+0.05}_{-0.04}$ \\
$\tau_p$ [keV] & -- & $7.07^{+0.06}_{-0.06}$ \\
norm$_{CompTT}$ & -- & $0.0228^{+0.0003}_{-0.0003}$ \\
E$_{gabs}\,$[keV] & $46.8^{+1.0}_{-0.8}$ & $48.0^{+0.8}_{-0.7}$ \\
$\sigma_{gabs}\,$[keV] & $7.7^{+1.0}_{-0.7}$ & $8.6^{+0.6}_{-0.5}$ \\
Strength${gabs}$ & $14.9^{+5.1}_{-2.6}$ & $21.3^{+2.0}_{-2.5}$ \\
Flux$^\dagger$& $1.499^{+0.003}_{-0.003}$ & $1.499^{+0.003}_{-0.003}$ \\
$\chi^2$/d.o.f. & $1645/1574$ & $1643/1573$ \\
\end{tabular}\end{ruledtabular}
\tablenotetext{}{
$^*$ In units of photons/keV/cm$^2$/s at 1 keV.\quad$^\dagger$ Flux calculated for the entire model in the $3-60\,$keV band and reported in units of $10^{-9}\,$erg\,cm$^{-2}\,$s$^{-1}$. Flux values with estimated errors were derived using the \texttt{cflux} model from \texttt{XSPEC} as resulting from FPMA.}
\end{table} 

\begin{figure*}[!t]
\centering
\includegraphics[width=.95\textwidth]{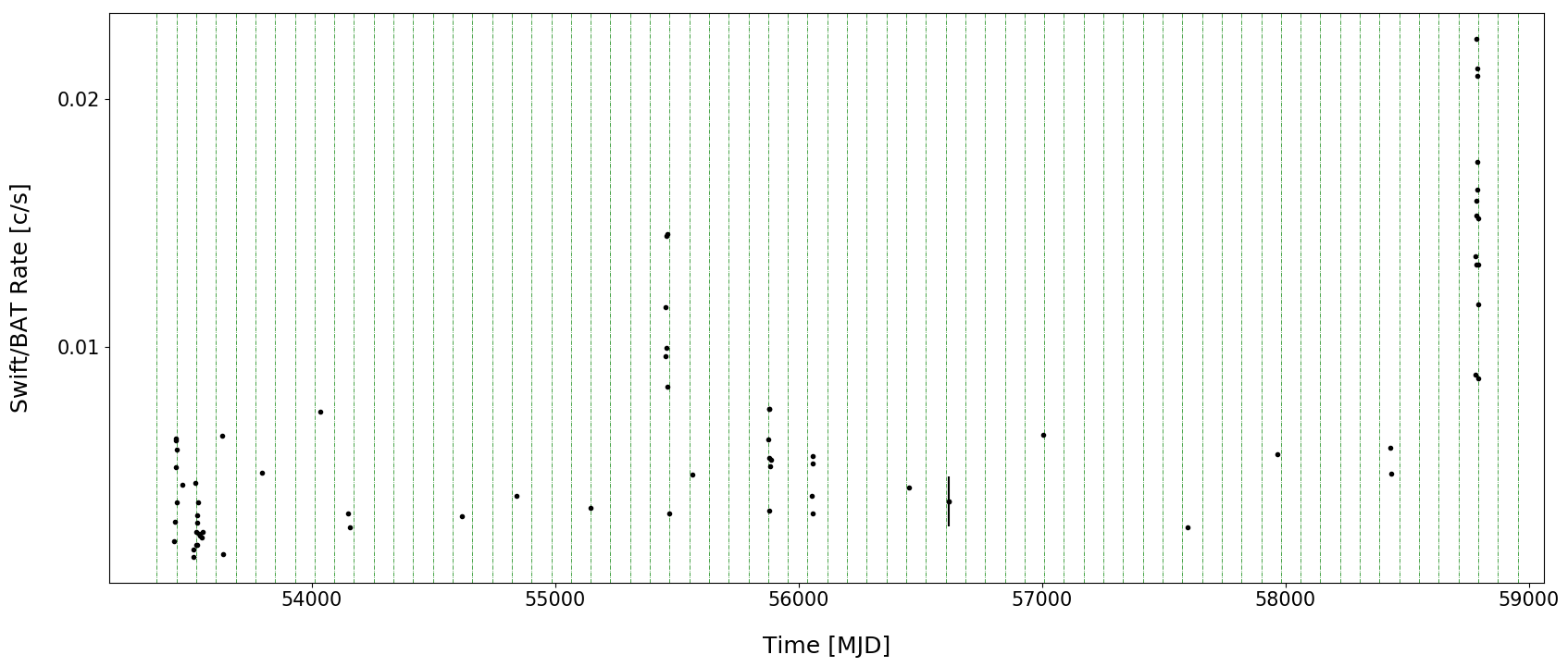}
\caption{\textit{Swift}/BAT light curve of XTE J1858+034. Only data points more significant than $3\sigma$ are reported. A typical error bar is shown only for one data point around MJD 56500 for clarity. Green dash-dotted lines are separated by 81 days. \label{fig:orbital_periods}}
\end{figure*}

To this aim, the publicly available spin-frequency values from GBM were used\footnote{\url{https://gammaray.nsstc.nasa.gov/gbm/science/pulsars/lightcurves/xtej1858.html}}.
The spin-up was measured during an interval of about 6 days around MJD 58786. 
The resulting spin-up value is $|\dot{P}|=10.5421(6)\,$s\,yr$^{-1}$ (see also \citealt[and references therein]{Malacaria20}).
Since the orbital parameters of this system are unknown, we tested the contribution of orbital modulation to the observed spin-up.
First, although \citet{Doroshenko08} report a possible orbital period for this source of about 380 days, we notice that its significance is low, while a visual inspection of the \textit{Swift}/BAT \citep{Krimm13} data\footnote{\url{https://swift.gsfc.nasa.gov/results/transients/weak/XTEJ1858p034/}} for this source revealed an outbursts recurrence of ${\sim}81$ days (see Fig.~\ref{fig:orbital_periods}), here assumed as the orbital period.
Moreover, for a K- or M-type optical companion star
(Tsygankov et al., in press), we adopted a value of the mass function $f(M)=M{_*}^3 sin^3i / (M{_* + M{_{NS}}})^2 = 1$, where $M{_*}$ and $M_{NS}$ are the mass of the companion and that of the NS, respectively, $i$ is the binary system inclination.
This corresponds to a value of the semi-major projected axis, $a_x\,sin\, i\simeq190\,$l-s.
Also, the data required only a small value of the eccentricity, assumed here as $e\simeq0.1$.
Finally, an epoch of $T0 =53436\,$MJD was chosen at the beginning of the first of the recurring outbursts.
An argument of periapse of $\omega=107^\circ$ was found to best-fit the GBM frequency values assuming no accretion torque.
This orbit assumed, the maximum orbital contribution to the spin-up was found to be only about $10\%$.

\begin{table}[!t]
\caption{Best-fit results of XTE J1858+034 spectral analysis with different configurations of the physical Bulk+Thermal Comptonization model \texttt{bwcyc}. 
All sets have the distance value fixed at $d=10.9\,$kpc.
All reported errors are at $1\sigma\,$c.l.} \label{table:spectral2}
\begin{ruledtabular}\begin{tabular}{lccc}
  & BWCYCa & BWCYCb & BWCYCc\\
  & & & \\
C$_{FPMB}$  & $0.748^{+0.001}_{-0.001}$ & $0.748^{+0.001}_{-0.001}$ & $0.748^{+0.001}_{-0.001}$\\
N$_{\textrm{H}}$ [$10^{22}\,$cm$^{-2}$]  & $8.6^{+0.2}_{-0.2}$ & $8.4^{+0.2}_{-0.2}$ & $8.4^{+0.3}_{-0.8}$ \\
E$_{K\alpha}\,$[keV]  & $6.47^{+0.02}_{-0.02}$ & $6.48^{+0.02}_{-0.02}$ & $6.47^{+0.02}_{-0.02}$\\
$\sigma_{K\alpha}\,$[keV]  & $0.27^{+0.02}_{-0.02}$ & $0.28^{+0.02}_{-0.02}$ & $0.27^{+0.02}_{-0.02}$\\
norm$_{K\alpha}$ ($10^{-4}$)  & $5.5^{+0.4}_{-0.4}$ & $5.6^{+0.3}_{-0.3}$ & $5.6^{+0.4}_{-0.4}$\\
$\xi$  & $4.2^{+0.7}_{-1.7}$ & $3.2^{+0.8}_{-0.8}$ & $2.9^{+1.0}_{-0.6}$\\
$\delta$  &  $0.4^{+0.3}_{-0.1}$ & $0.6^{+0.3}_{-0.1}$ & $0.7^{+0.3}_{-0.2}$ \\
B [$10^{12}\,$G]  & $4.4^{+0.5}_{-0.3}$  & $5.4$ (fixed) & $5.4$ (fixed)\\
$\dot{M}$ [$10^{17}\,$g/s]  & $1.2$ (fixed) & $1.2$ (fixed) & $1.1^{+0.1}_{-0.1}$ \\
T$_e$ [keV]  & $5.6^{+0.2}_{-0.6}$ & $5.5^{+0.1}_{-0.5}$ & $5.2^{+0.2}_{-0.4}$ \\
r$_0$ [m]  & $73^{+9}_{-22}$ & $66^{+22}_{-12}$ & $49^{+29}_{-10}$\\
d [kpc]  & $10.9$ (fixed) & $10.9$ (fixed) & $10.9$ (fixed)\\
E$_{gabs}\,$[keV]  & $48.3^{+0.7}_{-0.7}$ & $48.6^{+0.5}_{-1.3}$ & $48.3^{+1.0}_{-1.1}$\\
$\sigma_{gabs}\,$[keV]  & $10.3^{+0.7}_{-2.1}$ & $9.6^{+1.2}_{-1.2}$ & $9.3^{+1.3}_{-0.6}$\\
Strength$_{gabs}$ &  $25.4^{+4.2}_{-6.4}$ & $27.6^{+3.1}_{-4.2}$ & $25.3^{+4.0}_{-3.9}$ \\
Flux$^\dagger$  & $1.499^{+0.003}_{-0.003}$ & $1.499^{+0.003}_{-0.002}$ & $1.499^{+0.003}_{-0.003}$\\
$\chi^2$/d.o.f.  & $1648/1574$ & $1648/1575$ & $1648/1574$\\
\end{tabular}\end{ruledtabular}
\tablenotetext{}{
$^{*}$ unconstrained.
\quad $^\dagger$ Flux calculated in the $3-60\,$keV band and reported in units of $10^{-9}\,$erg\,cm$^{-2}\,$s$^{-1}$. Flux values with estimated errors were derived using the \texttt{cflux} model from \texttt{XSPEC} as calculated for FPMA.}
\end{table} 

Assuming a magnetic field strength of $5.4{\times}10^{12}\,$G and adopting the \textit{NuSTAR} measured flux of $1.5{\times}10^{-9}\,$erg\,cm$^{-2}\,$s$^{-1}$ (see Table~\ref{table:spectral} and \ref{table:spectral2}), the standard accretion-disk torque theory \citep{GLb} can be used to infer the distance of the source according to the equation:
\begin{equation}\label{eq:torque}
\begin{split}
-\dot{P}=5\times10^{-5}\,\textrm{s\,yr}^{-1}\,M_{1.4}^{-3/7}\,R_{6}^{6/7}\,I_{45}^{-1} \\ \times\,n(\omega_s)\,\mu^{2/7}_{30} (P\,L^{3/7}_{37})^2
\end{split}
\end{equation}

where $M_{1.4}$ is the NS mass in units of 1.4\,M$_\odot$, $R_6$ is the NS radius in units of $10^6\,$cm, $I_{45}$ is the moment of inertia in units of $10^{45}\,$g\,cm$^2$, $\mu_{30}$ is the magnetic moment in units of $10^{30}\,$G\,cm$^3$, $n(\omega_s)\approx1.4$ is the dimensionless torque, $P$ is the spin period in seconds and $L_{37}$ is the bolometric luminosity in units of $10^{37}\,$erg\,s$^{-1}$.
For a measured $|\dot{P}|=10.5\,$s\,yr$^{-1}$, Eq.~\ref{eq:torque} allows to infer a distance of $d=10.9\pm1.0\,$kpc (estimated uncertainty at $1\sigma\,$c.l.).
This distance value is also independently confirmed by the analysis of the optical companion star as reported in the accompanying paper by Tsygankov et al. (in press),
and it was used to characterize different configurations of the \texttt{bwcyc} model. The corresponding mass accretion rate $\dot{M}=1.2\times10^{17}\,$g\,s$^{-1}$ was adopted altogether, derived assuming a luminosity $L=\eta \dot{M}c^2$, with efficiency $\eta=0.2$ \citep{Sibgatullin00}. The different configurations of the tested model are reported in Table~\ref{table:spectral2} include a set with the magnetic field strength as a free parameter (\texttt{BWCYCa}), one with the magnetic field strength fixed to $5.4{\times}10^{12}\,$G (\texttt{BWCYCb}), and one with a free $\dot{M}$ and a fixed magnetic field strength (\texttt{BWCYCc}).

\subsection{Timing analysis}

For the timing analysis, the \texttt{nuproducts} task was used to obtain light curves out of calibrated and cleaned events. 
These light curves were corrected for livetime, exposure and vignetting effects, and were extracted in the following energy bands: $3-10$, $10-20$, $20-30$, $30-40$, $40-60$ and $3-60\,$keV.

\begin{figure}[!t]
\includegraphics[width=0.48\textwidth]{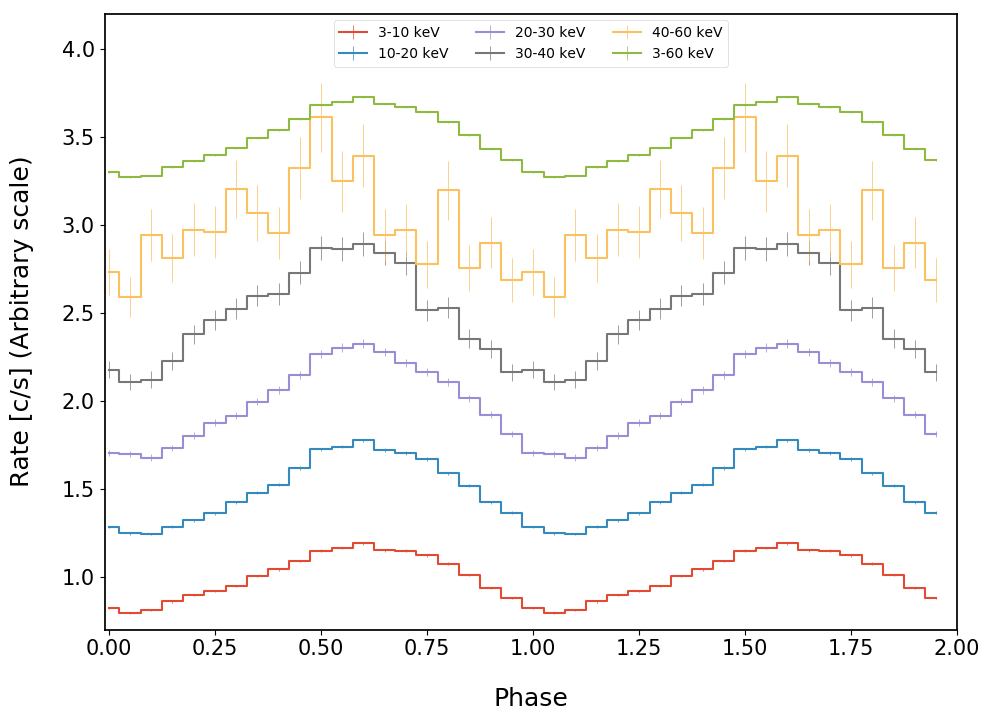}
\caption{XTE J1858+034 energy-resolved pulse profiles as observed by \textit{NuSTAR}.
The energy band increases upwards. Pulse profiles are shown twice in phase and rescaled in count rate for clarity.
\label{fig:pp}}
\end{figure}

All light curves were barycentered using the \texttt{barycorr} tool and the \textit{NuSTAR} clock correction file \texttt{nuCclock20100101v103}.
The light curve in the $3-60\,$keV energy band was binned to $5\,$s and used to search for pulsations around the known $221\,$s periodicity with the epoch folding method \citep{Leahy87}. 
The procedure results in a measured period of $P=218.393(2)\,$s. Pulsations are significant at $>99$\%. 
The uncertainty was estimated by simulating 500 light curves based on real data and altered with Poisson noise.

\begin{figure}[!t]
\includegraphics[width=0.48\textwidth]{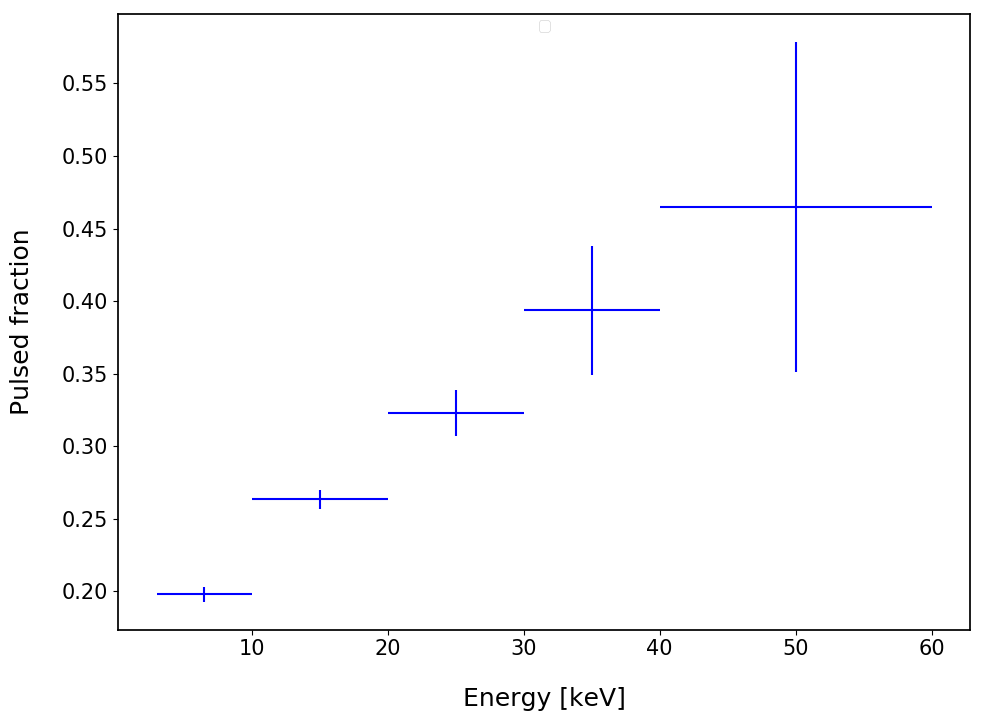}
\caption{XTE J1858+034 pulsed fraction as a function of the energy during the \textit{NuSTAR} observation in November 2019.
\label{fig:pf}}
\end{figure}

Light curves in different energy bands were folded to the best-fit spin period to obtain pulse profiles with a resolution of 20 phase bins (see Fig.~\ref{fig:pp}).
In turn, these were used to explore the pulsed fraction variation as a function of the energy (see Fig.~\ref{fig:pf}).
The pulsed fraction here is defined as $(I_{max}-I_{min})/(I_{max}+I_{min})$, where $I_{max}, I_{min}$ are the maximum and minimum pulse profile count rate, respectively.

\section{Discussion}

\subsection{Thermal Comptonization and a candidate cyclotron line}
The hard spectrum of XTE J1858+034 resembles that of other accreting X-ray pulsars observed both at low and high luminosity and well fit by a \texttt{CompTT} model \citep[and references therein]{Mukerjee20}.
The observation of \texttt{CompTT} spectra in accreting X-ray pulsars is usually interpreted as the result of thermal Comptonization processes in which the thermal energy of the accreting gas is transferred to the seed photons originating from the NS hotspots \citep{Becker+Wolff07}.
An increasing number of these sources also show that an additional \texttt{CompTT} component emerges in the high-energy range of the spectrum at low-luminosity stages \citep{2019MNRAS.487L..30T,2019MNRAS.483L.144T}. Although the formation of such component is not clear yet, it is likely due to a combination of cyclotron emission and following thermal Comptonized emission from a thin overheated layer of the NS atmosphere (see \citealt{2019MNRAS.483L.144T}, and references therein).
In this context, X Persei is a remarkable case since it has been shown that the cyclotron line in its spectrum can be mimicked by the convolution of the two \texttt{CompTT} spectral components around the energy where the flux from the low- and high-energy components is comparable \citep{Doroshenko12}.
However, among the sources whose spectrum is formed by two \texttt{CompTT} components, X Persei is the one with the highest electron temperature of the hard-energy \texttt{CompTT} component, $kT\sim15\,$keV.
If the absorption feature at $\sim48\,$keV in XTE J1858+034 is in fact resulting from the blend of two \texttt{CompTT} components, the high-energy \texttt{CompTT} would peak around $22\,$keV.
This would make XTE J1858+034 the most extreme among the X-ray pulsars that show such spectral shape.
However, such a spectral shape has so far only been observed in low-luminosity X-ray pulsars, while our analysis (see  Sect.~\ref{subsec:spectral}) 
show that the source is located at a relatively large distance of $10.9\,$kpc, thus implying a high-luminosity source (also supported by the analysis of Tsygankov et al., in press).
A second Gaussian or \texttt{CompTT} component that peaked above $45\,$keV was also tested in place of the absorption feature, but could not be successfully fit ($\chi^2_{red}>1.4$), although possibly due to the lack of statistics above $60\,$keV.

\subsection{Thermal and bulk Comptonization}

With the newly inferred source distance value of $d=10.9\,$kpc, the derived flux of $1.5\times10^{-9}\,$erg\,cm$^{-2}\,$s$^{-1}$ implies a luminosity of $2.1\times10^{37}\,$erg\,s$^{-1}$.
Adopting this distance value, all different sets of the \texttt{bwcyc} model are statistically equivalent and all parameters show acceptable values.
For the \texttt{BWCYCa} configuration,
the returned magnetic field strength of $4.2\times10^{12}\,$G is consistent within $2\sigma$ with that inferred from the candidate CRSF.
Notably, the \texttt{BWCYCb} model with fixed values of the distance, magnetic field strength and accretion rate also fits the data and returns acceptable values of the best-fit parameters.
The \texttt{BWCYCc} fit returns the smallest (but still acceptable) $r_0$ value, and a mass accretion rate $\dot{M}$ that is almost coincident with that inferred from the X-ray (isotropic) luminosity.

In any case, when interpreting the results from the \texttt{bwcyc} model, it is important to keep in mind that, as reported in \citet{Ferrigno09}, the \texttt{BWcyc} model may need adjustments in the spectral parameters with respect to the original prescriptions.
For example, the best-fit magnetic field value may differ from that inferred by the CRSF if the spectrum is formed at a NS site that is spatially different than the CRSF forming region. Likewise, the best-fit mass accretion rate $\dot{M}$ can be different from that inferred by the X-ray luminosity due to an uncertain efficiency conversion factor ($\eta$) and anisotropic emission.

\subsection{Timing results}

Pulse profiles of XTE J1858+034 as observed by\textit{NuSTAR} show a single-peak structure and a shape that is only weakly energy-dependent (see Fig.~\ref{fig:pp}).
This is typically observed at low mass accretion rates (see, e.g. \citealt{Malacaria+15}), and qualitatively interpreted as the beaming pattern resulting from a pencil-beam emission.
However, single-peaked pulse profiles are also observed at high accretion rates, like in the case of Pulsating Ultra-Luminous X-ray sources (PULXs), e.g. Swift J0243.6+6124, where single-peak pulse profiles persist at high luminosity and only switch to more complex profiles at super-Eddington luminosity \citep{Wilson-Hodge18}.

The pulsed fraction shows a considerable energy-dependence, and almost doubles from $20\%$ in the $3-10\,$keV to about $40\%$ in the $30-40\,$keV energy band, above which the lack of statistics prevent us from drawing firm conclusions.

\section{Conclusions}

We analyzed the \textit{NuSTAR} observation of the 2019 outburst of the XRB XTE J1858+034.
The source, relatively poorly studied, has now been characterized in multiple ways. A candidate cyclotron line is found in its spectrum at $48\,$keV.
This implies a magnetic field strength of $5.4\times10^{12}\,$G, consistent with the value obtained from the physical fitting model of thermal and bulk Comptonization \texttt{bwcyc} in its best-fit configurations.
We propose an orbital period of about 81 days based on the visual inspection of the X-ray outbursts recurrence time.
Arguments are given to review the previously proposed optical counterpart and its distance value in favor of a distance of $10.9\pm1.0\,$kpc obtained from standard accretion-torque theory.

\acknowledgments
\balance

This research has made use of data and software provided by the High Energy Astrophysics Science Archive Research Center (HEASARC), which is a service of the Astrophysics Science Division at NASA/GSFC and the High Energy Astrophysics Division of the Smithsonian Astrophysical Observatory. We acknowledge extensive use of the NASA Abstract Database Service (ADS). C.M. is supported by an appointment to the NASA Postdoctoral Program at the Marshall Space Flight Center, administered by Universities Space Research Association under contract with NASA. AAL and SST acknowledge support from the Russian Science Foundation (grant 19-12-00423).

\bibliographystyle{yahapj}
\bibliography{references}
\end{document}